 \documentclass[CEJP,PDF]{cej} 
\usepackage{layout}
\usepackage[OT4]{fontenc}
\usepackage{amssymb}
\usepackage{amsmath}
\usepackage{amsfonts}
\usepackage{graphics}
\usepackage{graphicx}
\usepackage{epstopdf}
\usepackage{textcomp}
\usepackage{hyperref}

\newcommand{\twopartdef}[4]
{
	\left\{
		\begin{array}{ll}
			#1 & \mbox{if } #2 \\
			#3 & \mbox{if } #4
		\end{array}
	\right.
}

\title{New orderings for effective mass Hamiltonians and associated wavefunction matching conditions at abrupt heterojunctions}

\articletype{Editorial}

\author{M. Vubangsi \inst{1}\email{vmercel@yahoo.com},
        M. Tchoffo\inst{1}\inst{2}\email{mtchoffo2000@yahoo.fr},
        L. C. Fai \inst{1}\inst{2}\email{corneliusfai@yahoo.com}}

\institute{
     \inst{1} Mesoscopic and Multilayer Structures Laboratory,
     Department of Physics, Faculty of Science, \\
     University of Dschang, P.O. Box 67 Dschang, Cameroon
     \inst{2} Department of Physics, H.T.T.C. Bambili, University of Bamenda, P.O. Box 39, North West Region, Cameroon
          }

\abstract{We show that for a given setting of the Von Roos ambiguity parameters, it is not required that the Morrow and Brownstein condition be satisfied to achieve physically acceptable transport properties at abrupt heterojunctions. We have derived two such settings and corresponding envelope function matching conditions.
}


\begin{document}
\maketitle


\section{\label{intro} Introduction}

The most general form of the kinetic energy operator for a system endowed with a coordinate dependent mass was first proposed by O. von Roos \cite{1} as follows:
\begin{equation}
\label{eq.1}
\hat{T} =\frac{1}{4}\left(m^{\alpha}\mathbf{p}m^{\beta} \mathbf{p}m^{\gamma} + m^{\gamma}\mathbf{p} m^{\beta} \mathbf{p}m^{\alpha} \right) 
\end{equation}
where $m = m(x)$ is the position dependent effective mass (PDEM) and $\mathbf{p}$ the momentum operator. The arbitrary constants $\alpha, \beta$ and $\gamma$ are known as the Von Roos ambiguity parameters. They are linked by the following constraint: 
\begin{equation}
\label{eq.2}
\alpha + \beta + \gamma = -1
\end{equation}
Eq.\ref{eq.1} has been extensively applied in the study of variable mass systems \cite{2,3,4,5,6}. Such studies are particularly of import in the search for electronic properties of semiconductor nano structures \cite{7,8,9,10}. Despite the prominence of eq.~(\ref{eq.1}) in the literature, the recipe of settings of the Von Roos parameters for acceptable kinetic energy operators is rather limited, at least basing on some validation rules such as by Dutra and Almeida \cite{11},  Mustafa and Mazharimousavi \cite{12} and Gonul et al. \cite{13}

In \cite{14}, it was obtained that for admissible continuity conditions at the heterojunction boundaries between two crystals, the following condition is requisite:
\begin{equation}
\label{eq.3}
\alpha = \gamma
\end{equation}
implying that the constraint on the ambiguity parameters reduces to:

\begin{equation}
\label{eq.4}
2\alpha + \beta = -1
\end{equation}

This work has a double purpose. First, we wish to re-examine condition eq.~(\ref{eq.3}) with respect to non-vanishing envelope functions at abrupt heterojunctions and then to propose a new set of orderings for effective mass kinetic energy operators.

The paper is organized as follows: In section \ref{sec1}, We determine a set of orderings for which the time-independent Schr\"odinger equation with eq.~(\ref{eq.1}) as kinetic energy operator is mapped to an isospectral constant mass problem. In section \ref{sec2}, We determine the envelope function continuity conditions and apply these to an abrupt heterojunction with step potential and step mass. The results are discussed in section \ref{sec3} and we end with concluding remarks in section \ref{sec4}.

\section{\label{sec1} Settings for the ambiguity parameters}

Let's consider the 1D time-independent Schr\"odinger equation for a particle with PDEM:

\begin{equation}
\label{eq.5}
\hat{T}\psi(x)+V(x)\psi(x) = E \psi(x)
\end{equation}
By putting the momenta in the kinetic energy operator to the right and working with $\hbar^{2}=2$, one obtains:

\begin{equation}
\label{eq.6}
\frac{\mathrm{d}^{2}}{\mathrm{d}x^{2}}\psi(x) - \frac{m'}{m}\frac{\mathrm{d}}{\mathrm{d}x}\psi(x) + \left[ \eta_{1}\frac{m^{\prime 2}}{m^{2}} +\eta_{2}\frac{m^{\prime \prime}}{m}+\frac{2m}{\hbar^{2}}\left(E-V(x)\right)\right] \psi(x) =0
\end{equation}
the primes represent derivation with respect to $x$, and $\eta_{1,2}$ are defined as:

\begin{equation}
\label{eq.7}
\eta_{1} = \frac{1}{2}\left( \alpha^{2}+\gamma^{2} + \alpha\beta+\gamma\beta -\alpha-\gamma\right) 
\end{equation}
\begin{equation}
\label{eq.8}
\eta_{2} = \frac{1}{2}\left(\alpha+\gamma \right) 
\end{equation}
We perform the following coordinate transformation as per the Sturm-Liouville approach:

\begin{eqnarray}
\label{eq.9}
y(x) &&= \int_{}^{x}\sqrt[]{m}\mathrm{d}x \nonumber \\
\phi(y) &&= \sqrt[4]{m}\psi(x)
\end{eqnarray}
As such, eq.~(\ref{eq.6}) takes the form:

\begin{equation}
\label{eq.10}
\frac{\mathrm{d}^{2}}{\mathrm{d}y^{2}}\phi(y) + \left[E - V(x) +\left(\eta_{1}-\frac{7}{16} \right)\frac{m^{\prime 2}}{m^{3}}+\left(\eta_{2}+\frac{1}{4} \right)\frac{m^{\prime \prime}}{m^{2}}   \right]\phi(y) = 0  
\end{equation}

To suppress in eq.~(\ref{eq.10}) the dependence on the ambiguity parameters, we impose the following two conditions:

\begin{equation}
\label{eq.11}
\eta_{1}-\frac{7}{16}=0
\end{equation}
and
\begin{equation}
\label{eq.12}
\eta_{2}+\frac{1}{4} = 0
\end{equation}
In this way, we obtain eq.~(\ref{eq.10}) as:

\begin{equation}
\label{eq.13}
\frac{\mathrm{d}^{2}}{\mathrm{d}y^{2}}\phi(y) + \left[E - V(x) \right]\phi(y) = 0  
\end{equation}
The target system constructed eq.~(\ref{eq.13}) is isospectral and isopotential to the initial one \ref{eq.5}. The difference between the two lies only in the quantum states. This is a unique feature that differentiates this work from the rest in the literature.

At the heterojunction boundary, it is required that the following quantities be continuous:

\begin{eqnarray}
\label{eq.14}
\phi(y) \qquad \text{and} \qquad \frac{\mathrm{d}}{\mathrm{d}y}\phi(y)
\end{eqnarray}
Translating into $x-$space, continuity is imposed on:
\begin{eqnarray}
\label{eq.15}
\frac{1}{\sqrt[4]{m(x)}}\phi(y(x)) \qquad \text{and} \qquad \frac{1}{m(x)^{1/2}}\frac{\mathrm{d}}{\mathrm{d}x}\frac{1}{\sqrt[4]{m(x)}}\phi(y(x))
\end{eqnarray}

The equations eq.~(\ref{eq.11}) and eq.~(\ref{eq.12}) are two constraints that should yield the settings for the ambiguity parameters for which the new constant mass problem eq.~(\ref{eq.13}) has exactly the same potential and energy spectrum as the original problem. To determine these settings, we start by substituting eq.~(\ref{eq.2}) in eq.~(\ref{eq.12}) and we obtain:

\begin{equation}
\label{eq.16}
\beta = -\frac{1}{2}
\end{equation}
With this value of $\beta$ it is straight forward to obtain from eq.~(\ref{eq.11}) and eq.~(\ref{eq.12}) the following relations:

\begin{equation}
\label{eq.17}
16\gamma^{2}-8\gamma-1=0, \qquad \alpha=-\frac{1}{2}-\gamma
\end{equation}

These results give rise to new orderings as shown on table \ref{table1}.

\renewcommand{\arraystretch}{2.0}
\begin{table}[h]
\caption{New sets of orderings}
\label{table1}
\begin{center}
\begin{tabular}{|c|c|c|c|}
\hline
 & $\alpha$ &     $\beta$     & $\gamma$ \\ 
 \hline \hline
 Set 1: & $-\frac{1}{4}\left(3-\sqrt{2}\right)$ &     $-\frac{1}{2}    $ & $\frac{1}{4}\left(1-\sqrt{2}\right)$ \\ 
 \hline
 Set 2: & $-\frac{1}{4}\left(3+\sqrt{2}\right)$ & $    -\frac{1}{2}$     & $\frac{1}{4}\left(1+\sqrt{2}\right)$ \\ 
 \hline
\end{tabular} 
\end{center}
\end{table}

The results on Table \ref{table1} conform to the Von Roos constraint eq.~(\ref{eq.2}) and both orderings a admissible according to \cite{11,12,13}. Based on the orderings in table \ref{table1}, the distinctive feature between \ref{eq.1} and other results in the literature is that the effective potential here is apriori not mass dependent which renders application of results of the model in actual physical systems possible.

\section{\label{sec2}Transport properties at abrupt heterojunctions}

We model the system with the potential and mass function having a first order discontinuity at $x=0$:
\begin{equation}
\label{eq.18}
V(x) = V_{0}\Theta(x)
\end{equation}
\begin{equation}
\label{eq.19}
m(x) = m_{1}+\left(m_{2}-m_{1}\right)\Theta(x)
\end{equation}
with 

\begin{equation}
\label{eq.20}
\Theta(x) = \twopartdef{0}{x<0}{1}{x>0}
\end{equation}

Using \ref{eq.1}, the continuity conditions at $x=0$ are written as follows:

\begin{eqnarray}
\label{eq.21}
\frac{1}{\sqrt[4]{m_{1}}}\phi^{-}(y^{-}(0)) &&=\frac{1}{\sqrt[2]{m_{2}}}\phi^{+}(y^{+}(0)) \nonumber \\
\frac{1}{m_{1}^{3/4}}\frac{\mathrm{d}}{\mathrm{d}x}\phi^{-}(y^{-}(0)) &&= \frac{1}{m_{2}^{3/4}}\frac{\mathrm{d}}{\mathrm{d}x}\phi^{+}(y^{+}(0))
\end{eqnarray}
The superscripts $-,+$ denote solutions in to the left and right of $x=0$ respectively.
where;

\begin{equation}
\label{eq.22}
\frac{1}{\sqrt[4]{m(x)}}\phi(y(x)) = \psi(x)= \twopartdef{\frac{1}{\sqrt[4]{m_{1}}} \exp{\left(\mathrm{i}k_{1}x\right)} + \mathrm{r}\frac{1}{\sqrt[4]{m_{1}}} \exp{\left(-\mathrm{i}k_{1}x\right)}}{x<0}{\mathrm{t}\frac{1}{\sqrt[4]{m_{2}}} \exp{\left(\mathrm{i}k_{2}x\right)}}{x>0}
\end{equation}
with $k_{1}=\sqrt{m_{1}E}$ , $k_{2}=\sqrt{m_{2}(E-V_{0})}$ and $\mathrm{r}$ and $\mathrm{t}$ are the reflection and transmission amplitudes respectively

The connection rules at $x=0$ give:

\begin{eqnarray}
\label{eq.23}
\frac{1}{\sqrt[4]{m_{1}}}(1+\mathrm{r})&&=\frac{1}{\sqrt[4]{m_{2}}}\mathrm{t} \nonumber \\
\frac{1}{m_{1}^{3/4}}(1-\mathrm{r})&&=\frac{1}{m_{2}^{3/4}}k_{2}\mathrm{t}
\end{eqnarray}
whereof the reflectivity $R$ and transmittivity $T$ are obtained:
\begin{equation}
\label{eq.24}
R = \left|\frac{\sqrt{E-V_{0}}-\sqrt{E}}{\sqrt{E-V_{0}}+\sqrt{E}} \right|^{2} 
\end{equation}
\begin{equation}
\label{eq.25}
T = \frac{4\alpha\left|\sqrt{E(E-V_{0})} \right| }{\left|\sqrt{E}+\sqrt{E-V_{0}} \right|^{2} }
\end{equation}
being $\alpha=m_{2}/m_{1}$ the mass discontinuity. In the limit $E>>V_{0}$ The transparency of the barrier remains sensitive to the mass discontinuity as then we have:

\begin{equation}
\label{eq.26}
T\approx \frac{m_{2}}{m_{1}}
\end{equation}

\section{\label{sec3} Discussion of Results}
Formulating the correct Hamiltonian for a particle with spatially dependent mass in an arbitrary potential well has for a long time drawn much attention. Due to the fact that the effective mass and the momentum operator no longer commute, ordering these in the kinetic energy operator becomes nontrivial.  

By imposing $\alpha=\gamma$ though, it turns out that $\beta=-1/2$ and $\alpha=\gamma=-1/4$ which is the ordering that yields the correspondence between the classical and quantum mechanical PDEM system \cite{15}. The same ordering is derived in \cite{16} with the assumption:
\begin{equation}
\label{eq.27}
\lim\limits_{E\rightarrow \infty}T=1
\end{equation}
In contrast to this, our result eq.~(\ref{eq.26}) shows that the transmittivity is unlikely to follow the trend of a constant mass system when the particle is endowed with sufficiently high energy above the barrier height. 

\section{\label{sec4} Conclusion}
We have shown that even for $\alpha \neq \gamma$ feasible transport properties are obtainable at an abrupt heterojunction for quantum systems with PDEM. 

We have supplemented the recipe of orderings in the literature with two new settings which are verified to be admissible.




\end{document}